# Multiferroic interfaces composed of $d^0$ perovskites oxides

Yi Yang[1,2], Chen-Sheng Lin[1], Jin-Feng Chen[1,2], Lei Hu[1,2] and Wen-Dan Cheng[1]

[1] State Key Laboratory of Structural Chemistry, Fujian Institute of Research on the Structure of Matter, Chinese Academy of Sciences, Fuzhou, 350002, Fujian, People's Republic of China

[2] University of Chinese Academy of Sciences, Beijing, 100049, People's Republic of China

E-mail: cwd@fjirsm.ac.cn

**Abstract.** We investigate the electronic, ferroelectric and magnetic properties of $KTaO_3/PbTiO_3$ interfaces by using conventional density functional theory (DFT) and advanced DFT such as hybrid functional HSE06. We show that doped holes in valence bands or electrons in conduction bands give rise to ferromagnetism at the interfaces. The ferromagnetic states are ground states for both hole-doped (*p*-type) and electron-doped (*n*-type) interfaces by comparison with their corresponding nonmagnetic and antiferromagnetic states. Carriers (holes or electrons) concentrate near the interface to screen the polarization charge and thus the concentration of carrier varies with the ferroelectric polarization. Furthermore, the interface magnetization, which is nearly proportional to the concentration of carrier, can be tuned by ferroelectric polarization reversal, leading to strong intrinsic magnetoelectric effects at the interface of originally nonmagnetic $KTaO_3$ and $PbTiO_3$. Interestingly, a ferromagnetic-nonmagnetic transition tuned by an applied electric field can be realized at the *p*-type interface. This suggests an illuminating approach to multiferroic materials beyond conventional single-phase multiferroics and multi-phase multiferroics such as ferroelectric/ferromagnet heterostructures. The $KTaO_3/PbTiO_3$ interfaces may be promising in future multiferroic devices applications.

1. Introduction

Oxide interfaces are attracting a lot of interests not only for their crucial importance in technological applications of all-oxide devices, but also for their offering possible approaches to new, unexpected material properties and physics [1-3]. The typical LaAlO$_3$/SrTiO$_3$ interface is one of those interfaces. Ohtomo and Hwang found that the interface between two insulators displays a quasi-two-dimensional electron gas(2DEG) behaviour [4]. Furthermore, the interface of these two nonmagnetic perovskites also exhibit unexpected magnetic-behaviours[5]. This type of magnetic interface provides a platform to study new mechanisms of ferromagnetism in oxides. Electronic reconstruction occur at the interface between LaAlO$_3$ with alternating charged layers and SrTiO$_3$ with nominally neutral layers along (001) direction, which result in the occupancy of Ti-$3d$ orbitals and thus the conductivity and magnetism at the interface [2, 4, 6].

The observation of magnetism at LaAlO$_3$/SrTiO$_3$ interface stimulates interests in finding similar oxide heterostructures, which may exhibit ferromagnetism and conductivity [7, 8]. A(I)B(V)O$_3$-type (I and V indicate nominal charges of cations)perovskites with alternating charged layers along (001) direction have similar polar characteristic to A(III)B(III)O$_3$ such as LaAlO$_3$. Accordingly, the A(I)B(V)O$_3$ compound may be used as a component of a polar/non-polar interface for exploring magnetism and 2DEG. Based on density functional theory with a Hubbard $U$ (DFT+$U$) spin-polarized calculations, Oja et al. found that $p$-type NaNbO$_3$/SrTiO$_3$ interfaces (with doped holes) favor ferromagnetic (FM) solution in energy rather than antiferromagnetic(AFM) solution [9]. Recently, they demonstrated that ferromagnetism appear at $p$-type KTaO$_3$/SrTiO$_3$ interfaces by using density functional theory (DFT) calculations and experiments [7]. They suggested that $p$-type interface provides sufficient holes within the flat valence band, which results in high density of states (DOS) at the Fermi level ($D(E_f)$) and the satisfaction of the Stoner FM criterion $D(E_f)J > 1$, where

$J$ denotes the strength of exchange interaction.

However, ferromagnetism seems to be rare at *n*-type interfaces (with doped electrons) made of a A(I)B(V)O$_3$ type material and a non-polar A(II)B(IV)O$_3$ material. DFT+$U$ study shows that the ground state is nonmagnetic for small $U$ (smaller than 5 eV), while magnetic ordering favors AFM ordering rather than FM ordering for large $U$ (6~7 eV) at *n*-type KTaO$_3$/SrTiO$_3$ interfaces [7]. Again, the study by local spin density approximation (LSDA) with Hubbard $U$ calculations for *n*-type NaNbO$_3$/SrTiO$_3$ interfaces shows that the stability of FM solution in energy is close to that of AFM solution [9]. We believe that further studies are required to find possible robust ferromagnetism at the *n*-type interfaces.

For multiferroic devices applications, it is desirable to control magnetism by an applied electric field. The control of interface magnetization is possible by introducing ferroelectric (FE) component into polar/non-polar heterostructures, which obtain magnetization from electronic reconstruction at the interfaces. Our previous DFT study shows that it is possible to control interface magnetization by FE polarization reversal at the *p*-type interfaces between KTaO$_3$ and BaTiO$_3$ [8]. This results in a considerable magnetoelectric (ME) effect at the interfaces composed of nonmagnetic perovskites. For practical applications, it is more desirable that the ferromagnetic-nonmagnetic transition can be tuned by FE polarization switch, more than just a fraction of the total interface magnetization is changed during FE polarization reversal, as discussed in previous theoretical and experimental studies [8, 10-12], Therefore, more efforts are required to explore a typical system to achieve this aim. We choose KTaO$_3$ (KTO) and PbTiO$_3$ (PTO) as typical combinations to construct polar/non-polar interfaces. KTO is a quantum paraelectric material and become a FE within a biaxial in-plane epitaxial strain of SrTiO$_3$ substrate [13]. Calculated polarization under SrTiO$_3$ compressive strain is 0.313C/m$^2$ by using Berry phase theory [8, 14]. The strain-induced ferroelectricity of KTO may introduce a new tunable factor into heterostructure, and KTO can be a promising candidate for polar phase. For non-polar phase, we choose PTO as a constituent of heterostructure due to its large FE polarization. The measured remnant polarization for PTO films on SrTiO$_3$ substrate is about 0.72 C/m$^2$ [15]. The large polarization of PTO may produce large influence on interface properties. Recently, Pb(Zr$_{0.2}$Ti$_{0.8}$)O$_3$ (PZT: a famous variant of PTO) has been used in tuning conductivity of 2DEG due to its large spontaneous FE polarization [16].

In this study, we investigate the electronic, FE and magnetic properties of ideal defect-free KTO/PTO interfaces (both *p*-type and *n*-type) by using DFT methods (including advanced techniques such as hybrid functional HSE06), and prospect the ME effects at these interfaces. In section 2, we present the methods. We discuss the results in section 3, and finally make a conclusion in section 4.

## 2. Methods

We performed DFT calculations using projector augmented wave method within the Vienna ab initio simulation package (VASP) [17-19]. The plane-wave cutoff energy for all calculations is 500 eV. The semicore states of K, Ta, Pb, Ti are treated as valence electrons. Brillouin zone integrations are performed by mean of the tetrahedron method with Blöchl corrections [20] for electronic structure calculations, but for ionic relaxations, are performed using Gaussian smearing method with a broadening [21] of 0.1 eV. Monkhorst–Pack [22] k-point mesh is $8\times8\times N_1$ for ionic relaxation ($N_1$ changes accordingly based on the number of unit cells vertical to the interface). A denser k-point mesh of $15\times15\times N_2$ ($N_2$ changes the same way as $N_1$) is used in electronic structure calculations, except that mesh of $10\times10\times3$ is used in HSE06 calculations for $(KTO)_{1.5}/(PTO)_{1.5}$ superlattices. For expanded $p(2\times2)$ supercell calculations, mesh of $7\times7\times N_2$ are used. Both k-point mesh and cutoff energy result in good convergence of the computed ground-state properties.

In a heterostructure, the strain induced by lattice mismatch often has large influence on the structural and electronic properties, especially at the interface. For an accurate stimulation of the interface of FEs, it is necessary to reproduce the experimental lattices of bulk within DFT calculations. It is well known that the overestimated rate of the calculated lattice constants by the generalized gradient approximation (GGA) of Perdew-Burke-Ernzerhof (PBE) functional is close to the underestimated rate of those by the local density approximation (LDA). Recently, PBEsol, a revised version of GGA-PBE, has been proposed by Perdew and collaborators [23] to predict the solids and surfaces properties more accurately.

The optimized lattice constants of perovskites bulk SrTiO3 and KTO in cubic phases obtained with GGA-PBEsol are in better agreement with the experimental values than both the LDA and GGA-PBE values (see table 1). Furthermore, Lattice constants of PTO in tetragonal FE phase calculated with PBEsol are in excellent agreement with experimental values, while this is not the case for LDA or PBE. Our results are also consistent with previous calculations [24, 25]. Therefore, we use PBEsol as the exchange-correlation functional to stimulate KTO/PTO interfaces unless otherwise stated.

**Table 1.** Calculated lattice constants $a$ (Å) of cubic SrTiO3 and KTaO3, and the in-plane lattice constant $a$ (Å) and the tetragonality $c/a$ of the tetragonal PbTiO3 by using different density functionals (LDA, PBE and PBEsol). Experimental lattice constants and references are also presented.

|  | LDA | | PBE | | PBEsol | | Expt. | | References |
| --- | --- | --- | --- | --- | --- | --- | --- | --- | --- |
|  | a | c/a | a | c/a | a | c/a | a | c/a |  |
| SrTiO3 | 3.863 |  | 3.944 |  | 3.899 |  | 3.905 |  | [26] |
| KTaO3 | 3.960 |  | 4.030 |  | 3.992 |  | 3.984 |  | [27] |
| PbTiO3 | 3.865 | 1.045 | 3.847 | 1.231 | 3.880 | 1.076 | 3.88 | 1.071 | [28] |

It has been recognized that GGA underestimates the correlation effect in transition metal oxide, thus approaches beyond common DFT have to be considered. We utilize DFT including a Hubbard $U$ proposed by Dudarev et al. [29] to account Coulomb correlation effect. In this approach, effective $U_{eff} = U - J`$ are used, $U$ and $J`$ denote Coulomb electron-electron interaction. For *n*-type interfaces, $U_{eff} = 4\,eV$ is used for the fractional occupation Ti-*3d* orbitals [30, 31]. Nekrasov et al. [32] found that the use of correlation correction on oxygen 2*p* orbitals (by DFT+$U$ method) can significantly improve calculated spectral and magnetic properties of NiO, MnO, La2CuO4 in addition to the correlation correction on metal *d* orbitals. By using DFT+$U$ method, Pentcheva et al. [33] demonstrated that the intra-atomic electron coulomb repulsion on oxygen is essential to describe the character of hole localization on oxygen ions at *p*-type LaAlO$_3$/SrTiO$_3$ interfaces. At *p*-type KTaO$_3$/SrTiO$_3$ interfaces, Oja et al. [7] found that it is necessary to include on-site electron coulomb repulsion on oxygen *p* orbitals for describing magnetic properties and suggested a value of $U_{eff}=4\,eV$. We add coulomb correlation of $U_{eff}=4\,eV$ on the hole-doped *p*-orbitals in calculating electronic structures [7, 8]. The common DFT (without a Hubbard $U_{eff}$) is also employed to calculate electronic structures in order to make comparisons. Besides, Hubbard $U_{eff}$ value in range of 0 to 8 eV are used to show dependence of interface magnetic properties on correlation effect.

Besides using DFT+$U_{eff}$ capturing electron correlation, we also use the screened hybrid density functional of Heyd, Scuseria, and Ernzerhof (HSE06) [34] which can have an improved description of strongly electron correlated materials. The exchange functional of HSE06 is constructed by mixing 25% of the HF exchange with 75% of the PBE exchange. Previous studies show that HSE06-computed structural and electronic properties are significantly improved over conventional (semi)local functionals for transition metal oxide such as LaAlO$_3$, LaTiO$_3$, SrTiO$_3$ and PTO [35, 36]. Calculations for LaAlO$_3$/SrTiO$_3$ interfaces show that hybrid functional can effectively describe correlated electrons behavior and the energy splitting of $t_{2g}$ band at the conducting interface, while conventional DFT such as LDA is not the case [37]. For KTaO$_3$/SrTiO$_3$ interfaces, calculations show that hybrid functional can remedy the underestimation of correlation effects within LDA or GGA calculations [7]. Based on those findings, we believe that HSE06 can efficiently capture the electron correlation character of oxide interfaces.

We consider two types of interfaces, namely *p*-type with doped holes and *n*-type with doped electrons in KTO/PTO superlattices (figure 1) [38]. We use calculated SrTiO$_3$ lattice constant as the in-plane lattice constant of all superlattices to model epitaxial growth on the SrTiO$_3$ substrate. The interfaces are constructed in superlattices which composed of non-stoichiometric KTO and PTO within periodic boundary condition. All superlattices are optimized to the criterion that atomic forces are less than 0.01(eV/Å). In this situation, a superlattice contains the same type of two interfaces, *p*- or *n*-type. In paraelectric case, a

*p*-type (*n*-type) interface obtains half a hole (electron) charge per interface unit cell area, corresponding to the fully-compensated *p*-type (*n*-type) interface. Therefore, one hole (electron) resides in valence (conduction) bands in a single *p*-type (*n*-type) superlattice. In FE case, the two interfaces (with FE polarization pointing away or toward) are no longer the same. The reversal of FE polarization in one component makes the interchange of two interfaces. Particularly, the effect of FE polarization reversal on interface properties can be obtained by comparing properties of two interfaces in one superlattice. Using this concept, we discuss the appearance of ferromagnetic-nonmagnetic transition and strong ME effects.

## 3. Results and discussion

### 3.1. p-type interface

We consider short superlattice $(KTO)_{1.5}/(PTO)_{1.5}$ with two *p*-type interfaces. As seen from figure 2, under hole doping, the Fermi level moves into the valence band, resulting in possible metallicity. For spin-polarized case, there is an exchange splitting which indicates the FM character of interfaces. The coulomb correlation interaction enlarges the splitting and thus enhances magnetism at the interfaces (compare figure 2(b) with figure 2(a)). The interfaces display half-metallic FM property when a $U_{eff}$ of 4 eV is used. The electronic structure calculated by hybrid functional (HSE06) shows similar property, but with only a small minority DOS at the Fermi level (figure 2(c)). The energy gap between valence band maximum and conduction band minimum calculated in HSE06 methods is larger than that of DFT or DFT+$U_{eff}$, which is a well-known character of HSE06-calculated electronic structure, as demonstrated in previous studies [36, 39]. It is seen that the energy band of HSE06 spread wider as compared with DFT results. Orbital-projected Ti-3*d* and O-2*p* density of states show that Ti-3*d* and O-2*p* orbitals are main composition of the bands above the gap and the bands below the gap, respectively (see figure 2(c)). Besides, significant hybridization between Ti-3*d* and O-2*p* orbitals exists, as is shown by occupied Ti-3*d* states and unoccupied O-2*p* states. This leads to magnetic moments residing on Ti atoms for spin-polarized solution. Interestingly, the magnetic moments on Ti atoms have opposite direction to the adjacent O atoms(see figure 4(a)), similar to the results reported in $KTaO_3/BaTiO_3$ system [8]. Magnetic moments with opposite direction induced by hybridization also exist at Fe/BaTiO$_3$ interface, where hybridization between interface Ti and Fe atoms occur and result in negative magnetic moments on Ti atoms [10].

The origin of ferromagnetism at the hole doped interfaces can be understood in the Stoner model [40]. In the metallic band picture model, FM coupling occur when exchange interaction energy gain in spin-polarizing carriers has an advantage over kinetic energy increase, i.e. the satisfaction of the Stoner criterion $D(E_f)J > 1$. Excess holes emerge at the interface due to electronic reconstruction. Under hole doping, the Fermi level moves into the O-2p band, which has flat top in $d^0$ perovskites, resulting in large $D(E_f)$. This is confirmed by the fact that the Fermi level cross a peak for non-spin-polarized DOS, as seen from figure 2(a) and figure 2 (b). On the other hand, large $J$ of the O-2p orbitals also contributes to the appearance of ferromagnetism. It was found that $J$ of O-2p orbitals is larger than that of typical magnetic element Mn-3d orbitals [41].

The $p(1\times1)$ superlattice is expanded to the $p(2\times2)$ supercell to consider possible AFM configurations. Different AFM configurations are introduced by initiating specific magnetic moments arrangement, e.g. AFI state has alternate antiparallel spins along *a* direction and AFII state has antiparallel spins of checkerboard type in the interface plane (see insets of figure 3). Then, initial magnetic configurations are electronically optimized to stable states. The relative energy of three magnetic configurations to the nonmagnetic solution and the magnetic moments of FM states are shown in figure 3 as a function of on-site $U_{eff}$. Obviously, FM states are ground state for $U_{eff}$ from 0 to 8 eV. For typical $U_{eff}$, e.g. 4 eV, FM states obtain large stability in comparison with their corresponding AFM states. The holes are largely spin-polarized in DFT calculations and completely spin-polarized with a small $U_{eff}$. Thus, the large magnetization is another manifestation of the stability of the FM states. We note that the energy difference between FM states and nonmagnetic states in (KTO)/(PTO) here is close to that found for KTaO$_3$/SrTiO$_3$ system (compare figure 3 here with figure 2(a) in Ref. [7] ), where FM ordering temperatures in experiments are found to be well above room temperature. Since ordering temperature is proportional to the energy difference (or magnitude of exchange interaction) as a first approximation, the FM ordering temperature of KTO/PTO system may be higher than room temperature.

The effects of Coulomb correlation on inducing hole localization and charge disproportionation( hole charge prefer localizing in some oxygen ion rather than distributing evenly in interface AlO$_2$ plane) has been suggested in LaAlO$_3$/SrTiO$_3$ system [33]. In our case, we explore the influence of $U_{eff}$ on hole localization for FM and AFM configurations in the $p(2\times2)$ superlattice. We find that holes distribute evenly in four $p(1\times1)$ units of $p(2\times2)$ superlattice for FM configuration. Similarly, holes of AFM states delocalize just as the FM case, although opposite spins occupy different $p(1\times1)$ units in the ab plane. The calculations show that how holes spread in the superlattice is nearly independent of $U_{eff}$ and holes occupy O-2$p$ orbitals (Figure 4 shows the distribution of magnetic moments; the distribution of hole is similar to this due to the complete spin-polarization of holes). These holes prefer $p_\pi$ orbitals to minimize repulsion with positive charged cation, as suggested by electrostatic considerations (see figure 4(a)). The holes occupy the mixed oxygen orbitals of $p_xp_z$ or $p_yp_z$ in interface TiO$_2$ layers (figure 4(a)) and the $p_xp_y$ orbitals in PbO layer (figure 4(b)). The hole density in the superlattice show that hole doping mainly occur to PTO rather than KTO (figure 4(c)). With a moderate $U_{eff}$ of 4 eV, all magnetic states remain metallic as judged from their band structures.

We also consider longer $p(1\times1)$ and $p(2\times2)$ (KTO)$_{2.5}$/(PTO)$_{3.5}$ superlattices in which holes might spread over 3.5 unit cells of PTO. Metallic and FM properties are almost unchanged except that a small proportion of holes do not spin-polarize. The interface magnetization remains the same magnitude as that in short superlattice (KTO)$_{1.5}$/(PTO)$_{1.5}$. FM states remain ground state compared with AFM configurations.

To explore how holes and magnetization distribute in thick superlattices, we consider longer superlattice (KTO)$_{5.5}$/(PTO)$_{9.5}$. First, we focus on the FE properties of the optimized superlattice. Figure 5 shows the FE properties in the manner of relative displacement between cations (K,Ta, Pb and Ti) and oxygen anions in (KTO)$_{5.5}$/(PTO)$_{9.5}$ superlattice. It is seen that FE polarizations in the middle of PTO subunit are close to the polarization value in bulk PTO within the same SrTiO$_3$ strain. Large displacement difference between PTO and KTO subunits indicates that the spontaneous polarization of PTO is much larger than the polarization of KTO. In order to calculate the polarizations of the central PTO and KTO unit cells of the metallic heterostructure, new perovskites cells of five atoms are created with frozen structural relaxation corresponding to the unit cells in the heterostructure. By using Berry phase theory [14] implemented in VASP, the FE polarizations for middle PTO and KTO in the heterostructure are calculated to be 0.820 C/m2 and 0.356 C/m2, respectively. We then obtain the polarization charges of 0.44 e per interface unit cell area from the polarization difference. Large polarization difference results in negative and positive polarization charge at $I_L$ and $I_R$ interfaces, respectively. To reduce the depolarizing electric field, i.e. to decrease the electrostatic energy, holes collect at $I_L$ and reduce at $I_R$ to screen polarization charges. A quantitatively demonstration of the screening mechanism requires a comparison of screening hole charge with the polarization charge.

Next, we focus on the distribution of hole charge. Figure 6 (black short dash line) displays the distribution of hole along (001) direction. Holes concentrate at $I_L$ interface which the polarization of PTO points away from. Hole doping mainly occur at the two PTO cells next to $I_L$ and hole density decay fast away from $I_L$. Holes disappear at the $I_R$ interface which the polarization points to. The distribution of hole is consistent with polarizations difference of two constituents PTO and KTO. To relate the hole distribution with polarization charge, we calculate the screening charge at the interfaces.

In paraelectric case, where KTO and PTO are considered as paraelectric bulks, two interfaces are identical and thus hole density is the same at two interfaces, that is 0.5 e per interface unit cell area. This has been confirmed in previous study [8]. When the paraelectric superlattice is polarized to a FE one, the polarization charges show up at the interfaces due to the discontinuity of FE polarizations of two constituents. To screen the polarization charges, holes concentrate at $I_L$ which possesses negative polarization charges. In contrast, hole density decreases at the other interface. By integrating hole density in half of the superlattice around two interfaces, holes of $I_L$ and $I_R$ are 0.93 and 0.07e per unit cell, respectively. Therefore, the value of the screening charge is 0.43e per interface unit cell area (use the calculated hole density by the DFT plus $U_{eff} = 4\,eV$ method), which is about the same as the polarization charge (0.44e). Similar screening mechanism occur at *n*-type interfaces of $KNbO_3/PbTiO_3$ superlattice [42].

The interesting distribution of hole has tremendous effects on the spin-polarization of hole, and it leads to a considerable ME effect at the interface of nonmagnetic bulks. Figure 6 shows that the magnetic moment density has similar distribution to hole density. This is because hole doping enhances ferromagnetism. A large hole concentration at $I_L$ gives rise to a large local DOS at the Fermi level, which makes the satisfaction of the Stoner FM criterion. In this case the FM state is more favorable in energy than the nonmagnetic state [8]. This is consistent with the fact that the holes in two PTO unit cells at $I_L$ are largely spin-polarized while those in the PTO cells further away are partially or little spin-polarized (see figure 6(b)). The results of GGA are almost the same as those of GGA with a $U_{eff}$, except that coulomb correlation enhances ferromagnetism and gives rise to larger magnetization(Compare figure 6(b) to figure 6(a)).Therefore, large magnetization occurs at several PTO unit cells near $I_L$ while no magnetization exists at $I_R$. This suggests a ferromagnetic-nonmagnetic transition tuned by the reversal of electric polarization of PTO and results in a significant ME effect.

We calculate the interface magnetization by integrating magnetic moment density in half of the superlattice around the interfaces. The interface magnetization for $I_L$ and $I_R$ are 0.66 and -0.01 $\mu_B$ per unit cell area ($U_{eff}$ case). Therefore, the change in interface magnetization ($\Delta M$) caused by ferroelectric polarization switch is 0.67 $\mu_B$ per unit cell area. To estimate the ME effect, we adopt an approach proposed by Duan et al. [43], as has been used in our previous work [8]. The ME coefficient $\alpha_s$ is defined by $\mu_0 \Delta M = \alpha_s E$, where $\mu_0$ is the permeability of vacuum, $\Delta M$ is the induced change in interface magnetization by ferroelectric polarization switch (0.67 $\mu_B/a^2$, a=3.899 Å), and $E$ is the strength of applied electric field. Assuming an applied electric field of $250 \text{ kV/cm}$, a typical coercive field of PTO films in experiments [44, 45], we obtain ME coefficient $\alpha_S = 2.1 \times 10^{-10} \text{ G cm}^2/\text{V}$.

Comparing this coefficient with those predicted earlier for ferromagnet/ferroelectric interfaces, we find that the ME effect in our system is improved significantly. It is two orders of magnitude larger than the ME coefficient ($\sim 2.0 \times 10^{-12} \text{ G cm}^2/\text{V}$ (Ref.[43]) of the SrTiO$_3$/SrRuO$_3$ interface predicted by Rondinelli et al. [46]. In their study, the ME effect is ascribed to the spin-dependent screening of the polarization charges at the SrTiO$_3$/SrRuO$_3$ interface. By using the same electric field of $250 \text{ kV/cm}$, the ME coefficient for typical Fe/BaTiO$_3$ interface [10], is $\alpha_S = 1.2 \times 10^{-10} \text{ G cm}^2/\text{V}$. Thus the ME coefficient for interfaces of originally nonmagnetic bulk here is about twice as much as that for conventional ferromagnet/ferroelectric interfaces. This finding is inspiring in interface ME effect.

The electronic structures of KTO/PTO interfaces, composed of originally nonmagnetic bulks, reconstruct owing to polar discontinuity and give rise to the interface magnetization. Besides the ME coefficient, we also use a ratio of the change in magnetization to the total magnetization to consider the effect of ferroelectric reversal on the change of the interface magnetization. Table 2 lists our predicted result and other theoretical and experimental results for a comparison. As stated previously, a ferromagnetic-nonmagnetic transition (indicated by the ratio of nearly 100%) tuned by the reversal of electric polarization of PTO exists at the $p$-type KTO/PTO interface. In the typical Fe/BaTiO$_3$ heterostructure [10], adopting the largest change in interface magnetization and total magnetization of Fe/BaTiO$_3$ heterostructure by the DFT calculation there, the ratio is calculated to be about 1.2% (see table 2). In this heterostructure, the change of magnetization (resulting from Fe/BaTiO$_3$ interface bonding hybridization) is small by comparison with the magnetization of the Fe/BaTiO$_3$ heterostructure. Based on the experimental and simulation studies of Fe/BaTiO$_3$ heterostructures [11, 12], a new oxidized-type interface model (possesses an additional FeO monolayer between Fe and BaTiO$_3$ components) has been proposed. The new interface model was considered as the most probable model to describe the Fe/BaTiO$_3$ heterostructures in experiments [12]. It is difficult to measure the absolute value of magnetic moments on interface atoms by experiments. We adopt the magnetization change of DFT calculations and the total magnetization of the experimental Fe film to estimate the ratio [12].

As shown in table 2, the ME (or multiferroic) effect at $p$-type KTO/PTO interface is larger than those at conventional Fe/BaTiO$_3$ interfaces for several orders of magnitude. Moreover, two components of KTO/PTO interface are oxide perovskites, thus they have chemical and structural compatibility to each other, while Fe in Fe/BaTiO$_3$ heterostructure is easily oxidation. Therefore, *p*-type KTO/PTO interfaces may be excellent candidate for new type multiferroic and spintronic application.

**Table 2.** The ratio of the change in magnetization to the total magnetization at various multiferroic interfaces tuned by ferroelectric reversal (the calculating details are presented in text).

| Interfaces | *p*-type KTaO$_3$/PbTiO$_3$ | Fe/BaTiO$_3$ | Fe/FeO/BaTiO$_3$ |
|---|---|---|---|
| The ratio (%) | ~100 | ~1.2 | ~0.03 |
| Reference | Our study | [10] | [12] |

*3.2. n-type interface*

For *n*-type interfaces, doped electrons occupy the bottom of the conduction band. Figure 7(a) shows the spin-polarized and non-spin-polarized total DOS of *n*-type (KTO)$_{1.5}$/(PTO)$_{1.5}$ superlattice by using GGA+$U_{eff}$ method. It is seen that the doped electron occupies the conduction band, resulting in the metallicity of interfaces. Under common DFT, spin-polarized calculation results in a state with negligible magnetic moments, which is close to the non-spin-polarized state (its DOS is similar to the non-spin-polarized DOS of GGA+$U_{eff}$, thus it is not presented here). When $U_{eff}$ is switched on, a state with significant magnetization shows up, as indicated by the spin exchange-splitting in figure 7(a). Ti atoms of interface TiO$_2$ layer obtain significant magnetic moments. As seen from figure 7(b), the calculated electronic structures using HSE06 method are similar to the results of DFT+$U_{eff}$, except that larger gap and wider band are realized in hybrid functionals. The doped electron occupies the Ti-3*d* orbitals at the bottom of conduction band, as shown by the projected DOS in figure 7(b).

We use DFT+$U_{eff}$ method to find possible magnetic stable state based on the optimized structure from non-spin-polarized calculations. This is realized by expanding $p(1\times1)$

$(KTO)_{1.5}/(PTO)_{1.5}$ superlattice to the $p(2\times2)$ supercell to check whether the AFM states exist. We find no stable AFM configurations (magnetic moments on Ti atoms disappear or be negligible) by varying $U_{eff}$ from 0.0 to 7.0 eV. Figure 8 shows the stability of FM states in energy relative to their nonmagnetic states. It is seen that with moderate coulomb correlation interaction (e.g. $U_{eff} = 4\ eV$) FM states are the ground state. The large magnetization of the superlattice is consistent with the large stability of FM state. By projecting magnetization on local atoms, magnetic moments mainly focus on Ti atoms. For $p(2\times2)$ $(KTO)_{2.5}/(PTO)_{3.5}$ $n$-type superlattice, the doped electrons might spread widely. The calculated electron occupation shows that the charges spread over all Ti atoms, resulting in derivation from $d^1$ orbital occupation (also see figure 10). Therefore, metallic behavior remains at the interfaces. We find that FM properties are almost the same as those in $(KTO)_{1.5}/(PbTO)_{1.5}$.

To probe the FE properties and the distribution of electron density, we consider longer superlattice $(KTO)_{8.5}/(PTO)_{11.5}$. Figure 9 shows the FE properties in form of relative displacement between cations (K, Ta, Pb and Ti) and oxygen anions in $(KTO)_{8.5}/(PTO)_{11.5}$ superlattice. Similar to the $p$-type superlattice, the $n$-type superlattice also has large polarization difference between two subunits. This situation leads to negative and positive polarization charges at $I_L$ and $I_R$ interfaces, respectively. Therefore, the electrons reduce at $I_L$ and collect at $I_R$ to screen the polarization charges. Using the same method for $p$-type superlattice, the FE polarization for KTO and PTO are calculated to be 0.332 C/m² and 0.737 C/m², respectively. The polarization charge is 0.38 e per interface unit cell area. The electron distribution can be understood in carrier screening mechanism based on electrostatic consideration.

In order to calculate the screening charge, we calculate the distribution of electron density first (figure 10). As found in $(KTO)_{2.5}/(PTO)_{3.5}$ case, the electron resides mainly in PTO side. The electron charges of $I_L$ and $I_R$ are 0.16 e and 0.84 e per interface unit cell area by integrating electron density in half of the superlattice around two interfaces. Therefore, the screening charge is 0.34e per interface unit cell area, which is close to the polarization charge (0.38e).

As shown in figure 10, the doped electron resides mainly on PTO cells at $I_R$ and thus occupies the Ti-$3d$ orbitals, which may give rise to interface magnetization. By using spin-polarized calculations of both DFT and DFT with $U_{eff}$, we calculate magnetic moment density in $(KTO)_{8.5}/(PTO)_{11.5}$ $n$-type superlattice (figure 10). It is seen that PTO cells with large electron density (near $I_R$) are readily spin-polarized, especially within DFT+$U_{eff}$ calculations. Coulomb correlation interaction enhances ferromagnetism of Ti-3$d$ orbitals (see figures 10(a) and (b)). As is shown in figure 10, interface magnetization show up near $I_R$, while disappear at $I_L$. Then, a significant change of magnetization can be induced by FE polarization reversal. By integrating magnetic moment density in half of the superlattice around the interfaces, interface magnetizations are 0.09$\mu_B$ per interface unit cell area for $I_L$ and 0.82$\mu_B$ for $I_R$. Interface magnetization varies by 0.73 $\mu_B/a^2$ (a=3.899 Å) by FE polarization switch. Using the same method for $p$-type superlattice and assuming an applied electric field of $250\,kV/cm$, the ME coefficient is estimated to be $\alpha_S = 2.3\times10^{-10}\,G\,cm^2/V$. Therefore, the ME effect at the $n$-type interface is significant as that occur at $p$-type interface.

Comparing figure 9 with figure 10, we note that ferroelectricity coexists with metallicity at the interface. From common knowledge, these two properties may seem mutually incompatible with each other in the same phase. However, the recent studies[47, 48] show that FE metal phases of electron-doped BaTiO$_3$ are still retained up to a critical doped concentration, about 0.1 e per unit cell. Return to our results, we note that the PTO unit cells with large electron density have small ferroelectric displacement (see displacement and hole density at the $I_R$ in figure 9 and figure 10), while those with smaller electron density have displacement close to bulk( the value away from $I_R$ in figure 9 and figure 10). A large concentration of carrier leads to more effective screening of long range Coulomb interaction, which is responsible for the ferroelectric instability, stabilizing high-symmetry phase. Comparing figure 5 with figure 6, we find that the coexistence of ferroelectricity and metallicity also occur in $p$-type superlattice.

4. **Conclusions**

In conclusion, we have investigated the electronic, ferroelectric and magnetic properties of ideal defect-free KTO/PTO interfaces (both *p*-type and *n*-type) by using density functional theory methods (including advanced techniques such as HSE06). The results show that ME effects exist at the interfaces. Based on spin-polarized calculations, we find that spin-polarization is prone for both doped holes and electrons at the interfaces, and thus result in local magnetic moments on atoms near the interfaces. Further research for possible magnetic ground states shows that the FM states are ground states for both *n*-type and *p*-type interfaces by comparison with their nonmagnetic and AFM states. For *n*-type interface, doped electrons fill the Ti-3*d* orbitals and give rise to interface ferromagnetism. For *p*-type interface, doped holes occupy the O-2*p* band, resulting in $d^0$ ferromagnetism. Carriers (electrons or holes) concentrate at interfaces to screen the polarization charge and thus concentration of carriers can be regulated by ferroelectric polarization reversal. Furthermore, the interface magnetization, which is nearly proportional to the carrier density, can also be tuned by ferroelectric polarization, leading to strong magnetoelectric effects at the interface of originally nonmagnetic KTO and PTO. Our results suggest that KTO/PTO interfaces may be promising in multiferroic devices applications

Acknowledgments

The authors are grateful to Dr. Xianxin Wu in Institute of Physics, CAS and Dr. Chao He in Fujian Institute of Research on the Structure of Matter, CAS for valuable discussions. This investigation was based on work supported by the National Basic Research Program of China (2014CB845605), National Natural Science Foundation of China under Projects 21173225 and 91222204, and Fujian Provincial Key Laboratory of Theoretical and Computational Chemistry.

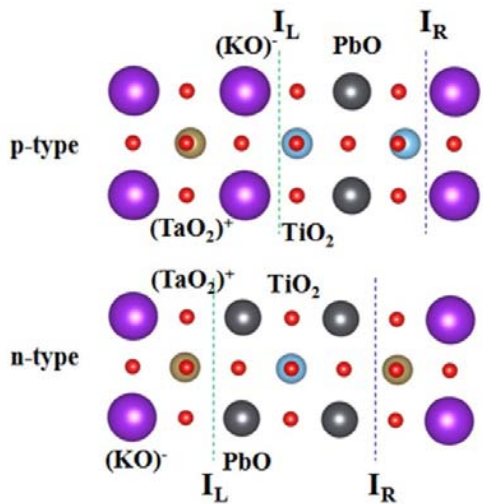

Figure 1. *p*-type and *n*-type (KTaO$_3$)$_{1.5}$/(PbTiO$_3$)$_{1.5}$ superlattices with the nominal ionic charges of charged layers. Longer superlattices are constructed by inserting several KTaO$_3$ and PbTiO$_3$ unit cells into the short superlattices. Ferroelectric polarization of PbTiO$_3$ subunit for all superlattices in present study is point from left interface (I$_L$) to right interface (I$_R$).

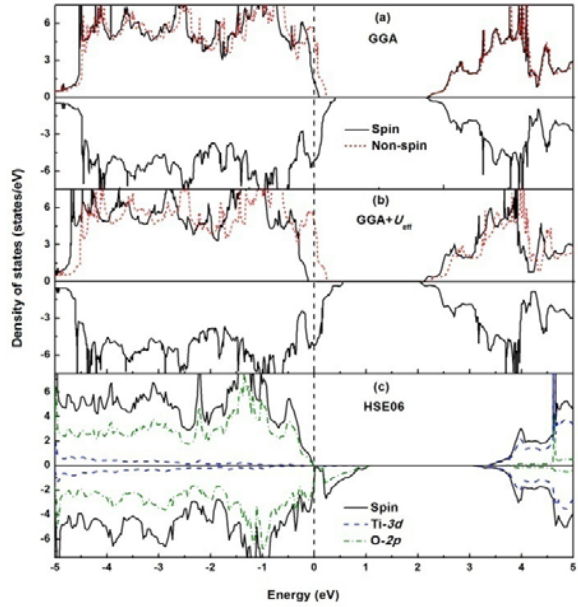

Figure 2. Spin-(black, solid line) and non-spin-polarized (red, short dash line) total density of states (DOS) for $(KTaO_3)_{1.5}/(PbTiO_3)_{1.5}$ $p$-type superlattice calculated by: (a) GGA and (b) GGA with $U_{eff} = 4\ eV$ on O-2$p$ orbitals; (c) Spin-polarized total DOS and projected DOS on Ti-3$d$ and O-2$p$ orbitals calculated by HSE06 method. Positive and negative DOS values denote the majority- and minority-spin, respectively. Vertical dash line indicates the Fermi level.

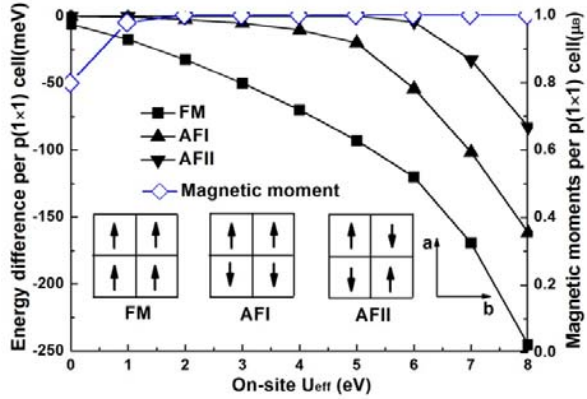

Figure 3. Energy difference between three magnetic states and the nonmagnetic state for $(KTaO_3)_{1.5}/(PbTiO_3)_{1.5}$ $p$-type superlattice as a function of correlation interaction $U_{eff}$; magnetization of ferromagnetic state (open blue diamond). Antiferromagnetic states are constructed in $a$-$b$ plane using chains($c$ direction) of parallel spins ($p(1\times1)$ as the basic unit). Antiferromagnetic-I (AFI) state(solid up triangle) has alternate antiparallel spin along $a$ direction. Antiferromagnetic-II(AFII) state(solid down triangle) has antiparallel spin of checkerboard type. Ferromagnetic state (solid square) has all parallel spin. The lines are guides for the eye.

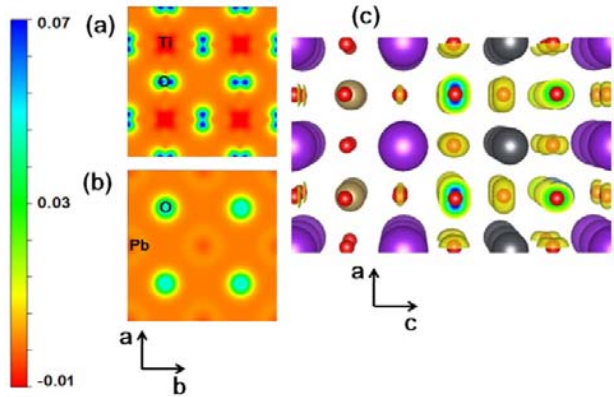

Figure 4. Magnetization(an interval of $0.01\mu_B a_0^{-3}$: $a_0$ is Bohr radius) of $(KTaO_3)_{1.5}/(PbTiO_3)_{1.5}$ $p$-type superlattice in (a) $TiO_2$ atomic plane at the left interface $I_L$, (b) PbO atomic plane, (c) whole superlattice.

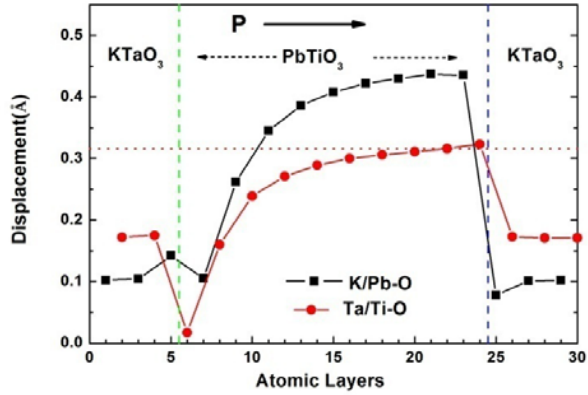

Figure 5. Ferroelectric displacements of Cations with respect to oxygen anions in $(KTaO_3)_{5.5}/(PbTiO_3)_{9.5}$ superlattice. The displacement is calculated in each atomic layer along the (001) direction. K-O or Pb-O(K/Pb-O) displacements are indicated by black square symbols. Ta-O or Ti-O(Ta/Ti-O) displacements are indicated by red circle symbols. The horizontal dot line indicates the Ti-O displacement in bulk $PbTiO_3$ within the same strain condition of the superlattice. The green and blue vertical dashed lines mark the left and right *p*-type interfaces, respectively. P with an arrow indicates the ferroelectric polarization direction in $PbTiO_3$ subunit.

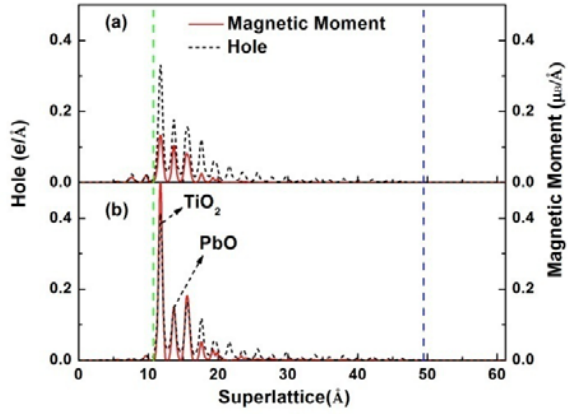

Figure 6. The hole density and magnetic moment density along (001) direction in $(KTaO_3)_{5.5}/(PbTiO_3)_{9.5}$ $p$-type superlattice for (a) GGA calculation, (b) GGA plus $U_{eff}$ (4eV) calculation. The integral of hole density along the superlattice is the doped hole in the valence band. The integral of magnetic moment density along the superlattice is the magnetic moment of the superlattice. $TiO_2$ (PbO) indicate hole density peak at the $TiO_2$ (PbO) layer. Vertical dash line at left and right denote the interfaces $I_L$ and $I_R$, respectively. Ferroelectric polarization direction in $PbTiO_3$ subunit is point from $I_L$ to $I_R$.

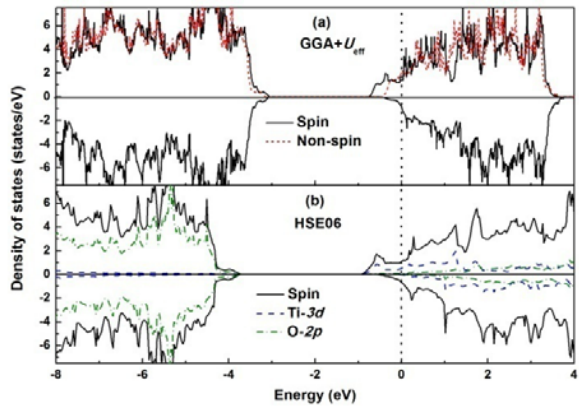

Figure 7. (a) Spin-(black, solid line) and non-spin-polarized (red, short dash line) total density of states (DOS) for $(KTaO_3)_{1.5}/(PbTiO_3)_{1.5}$ $n$-type superlattice calculated by GGA with $U_{eff}$ (4 eV) on Ti-$3d$ orbitals; (b) Spin-polarized total DOS and projected DOS on Ti-$3d$ and O-$2p$ orbitals calculated by HSE06 method. Positive and negative DOS values denote the majority- and minority-spin, respectively. Vertical dash line indicates the Fermi level.

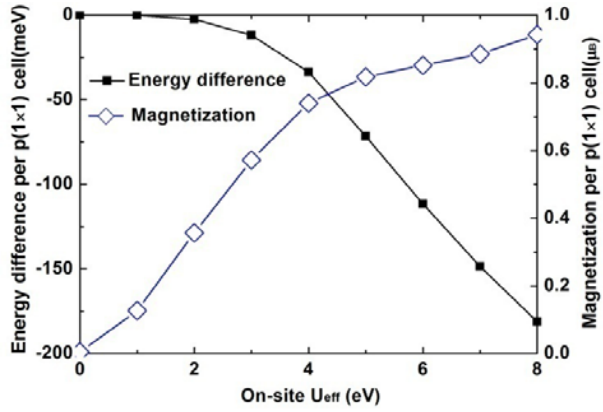

Figure 8. Energy difference between FM state and nonmagnetic state for $(KTaO_3)_{1.5}/(PbTiO_3)_{1.5}$ *n*-type superlattice as a function of correlation interaction $U_{eff}$; magnetization of FM state (open blue diamond).

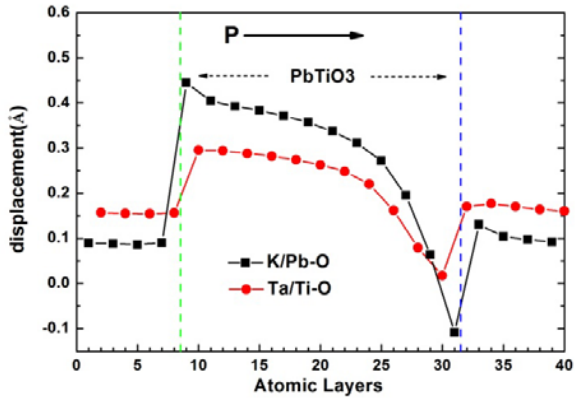

Figure 9. Ferroelectric displacements of cations with respect to oxygen anions in $(KTaO_3)_{8.5}/(PbTiO_3)_{11.5}$ superlattice. The displacement is calculated in each atomic layer along the (001) direction. K-O or Pb-O(K/Pb-O) displacements are indicated by black square symbols. Ta-O or Ti-O(Ta/Ti-O) displacements are indicated by red circle symbols. The horizontal dot line indicates the Ti-O displacement in bulk $PbTiO_3$ within the same $SrTiO_3$ strain condition of the superlattice. Vertical dash line at left and right denote the interfaces $I_L$ and $I_R$, respectively. P with an arrow indicates the ferroelectric polarization direction in $PbTiO_3$ subunit.

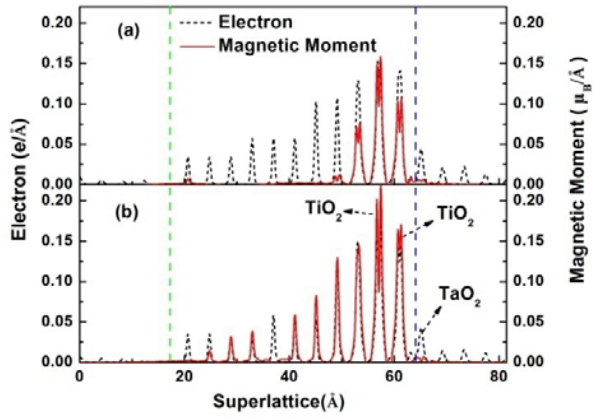

Figure 10. The electron density and magnetic moment density along the direction vertical to interface in $(KTaO_3)_{8.5}/(PbTiO_3)_{11.5}$ $n$-type superlattice for: (a) GGA calculation, (b) GGA+$U_{eff}$ (4eV) calculation. The integral of electron density along the superlattice is the electron in the conduction band. The integral of magnetic moment density along the superlattice is the magnetic moment of the superlattice. $TiO_2$ ($TaO_2$) indicate electron density peak at the $TiO_2$ ($TaO_2$) layer. Vertical dash line at left and right denote the interfaces $I_L$ and $I_R$, respectively. Ferroelectric polarization direction in $PbTiO_3$ subunit is point from $I_L$ to $I_R$.